\documentclass[11pt,a4paper]{article}
\usepackage[T1]{fontenc}
\usepackage[utf8]{inputenc}
\usepackage{lmodern}
\usepackage{amsmath,amssymb,amsthm,mathtools}
\usepackage{geometry,microtype,enumitem,booktabs,array}
\usepackage[colorlinks=true,linkcolor=blue!55!black,citecolor=blue!55!black,urlcolor=blue!55!black]{hyperref}
\usepackage{xcolor}
\setlist{itemsep=.15em,topsep=.3em}
\numberwithin{equation}{section}
\newtheorem{theorem}{Theorem}[section]
\newtheorem{proposition}[theorem]{Proposition}
\newtheorem{corollary}[theorem]{Corollary}
\theoremstyle{remark}\newtheorem{remark}[theorem]{Remark}
\newtheorem{example}[theorem]{Example}
\newcommand{\Tr}{\operatorname{Tr}}
\newcommand{\Prob}{\mathbb P}

\newcommand{\id}{\mathrm{id}}
\newcommand{\I}{\mathbb I}
\newcommand{\s}{\sigma}
\newcommand{\calA}{\mathcal A}
\newcommand{\comp}{\mathsf{Comp}}
\newcommand{\reality}{\mathsf E}
\newcommand{\predict}{\mathsf C}
\newcommand{\nodist}{\mathsf N}
\newcommand{\represent}{\mathsf K}
\newcommand{\intrinsic}{\mathsf R}
\title{\bfseries What Can Actually Be Proved About the\\ Einstein--Podolsky--Rosen Paradox?\\[.35em]\large Operational Consistency, Locality, and Quantum Completeness}
\author{Miko\l aj Sienicki\\
\small Polish--Japanese Academy of Information Technology\\
\small ul.~Koszykowa 86, 02-008 Warsaw, Poland, European Union
\and Krzysztof Sienicki\\
\small Chair of Theoretical Physics of Naturally Intelligent Systems (NIS)\\
\small ul.~Lipowa 2 / ul.~Topolowa 19\\
\small 05-807 Podkowa Le\'sna, Poland, European Union}
\date{11 November 2026 --- Revised research manuscript}
\begin{document}
\maketitle
\begin{abstract}
The Einstein--Podolsky--Rosen (EPR) argument distinguishes accurate quantum predictions from completeness of a physical description. Recent proposals have sought to settle EPR by appealing to certainty of prediction for correlated random outcomes. We separate predictive adequacy, operational description, the EPR criterion of reality, and ontological completeness. We prove perfect conditional predictability with uniform local marginals, no-signalling for unconditioned local instruments, and the CHSH exclusion of measurement-independent Bell-local models. We then give a setting-independent-seed, deterministic \emph{representation} of any finite conditional probability table and exhibit a Popescu--Rohrlich box to show why this representation by itself is neither a quantum model nor a deterministic physical dynamics. Most importantly, we formalize EPR's value- and context-indexed reality and completeness premises and construct a two-observable stochastic-source countermodel showing that the resulting incompleteness implication does not entail a claim excluding stochastic origin, while distinguishing this claim from fresh randomness at measurement. We derive a two-setting qubit steering bound to show that matching a restricted classical outcome table does not reproduce the corresponding trusted-qubit conditional-state assemblage. We distinguish the discrete-variable witness from covariance calculations for a \emph{stipulated} instantaneous, perfectly reflecting Gaussian collision map. Within that map we analyze mass-ratio dependence, incident spreading, illustrative mixed-state and detector noise, and the difference between fixed-axis Reid tests and optimized Gaussian steering. The resulting conditional thresholds are not estimates of experimental performance: the ion interaction, incident covariances, timing, and readout characteristics have not been independently established for the proposed apparatus. We also explain why an uncertainty relation alone does not establish intrinsic indeterminism. Finally, predetermined exponentially distributed lifetimes show that statistical memorylessness of radioactive decay alone cannot rule out predetermined event times; the scope of this counterexample is made explicit. None of these results establishes unrestricted quantum completeness or a new physical hidden-variable dynamics.
\end{abstract}

\section{Introduction}
Einstein, Podolsky and Rosen (EPR) asked whether the quantum-mechanical wave function gives a complete description of physical reality \cite{EPR1935}. Their argument combines perfect predictions for alternative measurements on distant systems with a sufficient criterion for an \emph{element of reality} and a premise that one can make the relevant inference without disturbing the distant system. It is not an allegation that the ordinary quantum probability rules are algebraically inconsistent.

Bell's theorem turns a precise family of assumptions about locality and independence into experimental constraints \cite{Bell1964,CHSH1969}. Quantum steering instead concerns whether the conditional states of one system admit a local-hidden-state model \cite{Wiseman2007,Uola2020}. Neither theorem, standing alone, proves that no conceivable ontologically richer description exists. Our earlier work on context-indexed elements of reality \cite{SienickiEPR} also informs the present analysis. This article substantially extends and supersedes the initial four-page comment published as \href{https://arxiv.org/abs/2604.13135v1}{arXiv:2604.13135v1} on 14 April 2026. Schnabel's revised preprint \cite{Schnabel2026} explicitly treats an incompleteness implication [I] and a predictability--randomness implication [II] as interchangeable, and invokes a collision-based entanglement preparation, an uncertainty relation, and an alpha-decay analogy. We assess the logical scope of those steps without claiming a novel discovery of standard Bell or probability theory.

The central contribution is an explicit separation between (a) the logical EPR premise, (b) operational random variables, (c) a convenient deterministic response-function representation, and (d) a physically specified deterministic ontology. This distinction lets us identify exactly which assumptions a claimed resolution of EPR must address.

\section{Four notions of completeness and the operational arena}
For a \emph{single-round} scenario \(\calA_1\), let \(s\) denote a preparation, \(x,y\) measurement settings, and \(a,b\) registered outcomes. Its single-round empirical content is
\begin{equation}
\mathcal P_{\calA_1}=\{p(a,b\mid x,y,s):(s,x,y)\in\calA_1\}.
\end{equation}
This table is not a complete characterization of general instruments or time-ordered experimental sequences. We distinguish:
\begin{enumerate}[label=(\roman*)]
\item \textbf{Predictive adequacy:} agreement of predicted distributions with a specified body of observations to stated accuracy.
\item \textbf{Operational completeness relative to an arena:} representation of all probabilities for experiments \emph{included in that arena}. Completeness relative to \(\calA_1\) only means adequacy for its single-round conditional table. For an extended arena \(\calA_{\mathrm{seq}}\), this must also encompass specified sequential interventions and their outcome probabilities.
\item \textbf{EPR completeness:} every element of reality identified using EPR's criterion has a counterpart in the theory \cite{EPR1935}.
\item \textbf{Ontological completeness:} no physically meaningful further specification is admissible beyond the theory's own state and dynamics. This statement makes sense only relative to an explicit class of admissible ontologies.
\end{enumerate}
These are different predicates. In particular, predictive adequacy does not imply ontological completeness, and rejecting Bell-local hidden variables does not reject all hidden-variable interpretations \cite{BohmI,BohmII}.

\begin{example}[Identical first-round statistics do not specify an instrument]
Let \(P_\pm=(\I\pm\s_z)/2\). Compare the L\"uders instrument
\begin{equation}
\mathcal I^{L}_{\pm}(\rho)=P_\pm\rho P_\pm
\end{equation}
with the measure-and-prepare instrument
\begin{equation}
\mathcal I^{R}_{\pm}(\rho)=\Tr(P_\pm\rho)\,|0\rangle\langle0|.
\end{equation}
For every input \(\rho\), both have outcome probabilities \(p(\pm)=\Tr(P_\pm\rho)\), and both sum to a trace-preserving completely positive map. For input \(|1\rangle\), however, the first instrument leaves the post-measurement state \(|1\rangle\), while the second resets it to \(|0\rangle\). A subsequent \(Z\) measurement therefore gives opposite certain outcomes. The two instruments are indistinguishable within the specified first-round probability table but distinguishable in a sequential arena \cite{DaviesLewis1970}.
\end{example}

\section{The original EPR structure}
In the continuous-variable idealization define
\begin{equation}
\widehat X=\hat x_A-\hat x_B,\qquad \widehat P=\hat p_A+\hat p_B.
\end{equation}
Canonical commutation relations imply
\begin{equation}
[\widehat X,\widehat P]=[\hat x_A,\hat p_A]-[\hat x_B,\hat p_B]=0,
\qquad [\hat x_B,\hat p_B]=i\hbar.
\end{equation}
The commuting collective combinations can be simultaneously sharp in an idealized EPR limit without making the individual position and momentum operators commute. The ideal limiting state is nonnormalizable; finite squeezing gives nonzero inferred variances and calls for operational criteria such as Reid's \cite{Reid1989}.

EPR's further inferential step asks what, under an assumption of no disturbance, may be attributed to system B when the experimenter at A could choose either position or momentum. The two incompatible alternatives are central to EPR's \emph{incompleteness} argument. A perfectly correlated single observable alone does not supply that simultaneous-attribution challenge, even though it can illustrate conditional probability.

\section{What quantum mechanics proves}
\subsection{Conditional certainty is compatible with non-degenerate marginals}
Let
\begin{equation}
|\Phi^+\rangle=\frac{|00\rangle+|11\rangle}{\sqrt2},\qquad \rho_{AB}=|\Phi^+\rangle\langle\Phi^+|.
\end{equation}
Write \(Z=\s_z\) and \(X=\s_x\).
\begin{theorem}[Perfect conditional prediction and random-looking marginals]
For either matched pair of measurements \(j\in\{X,Z\}\), Alice's and Bob's outcomes agree with probability one although each local outcome has probability \(1/2\):
\begin{equation}
\Prob(B_j=A_j\mid j,j)=1,\qquad
\Prob(B_j=+1\mid j)=\Prob(B_j=-1\mid j)=\tfrac12.
\end{equation}
The Bob observables \(X\) and \(Z\) are incompatible.
\end{theorem}
\begin{proof}
The reduced state is \(\rho_B=\Tr_A\rho_{AB}=\I_2/2\), giving both marginal distributions. The Bell state is a \(+1\) eigenstate of \(Z\otimes Z\) and \(X\otimes X\); matched outcomes therefore agree with certainty. Finally \([X,Z]=-2i\s_y\neq0\).
\end{proof}
This establishes compatibility of \emph{unconditional} uncertainty and \emph{conditional} predictability. The probabilities do not independently decide whether the source of outcomes is ontologically stochastic.

\subsection{No-signalling under a local instrument}
For a local quantum instrument \(\{\mathcal I_{a\mid x}\}_a\) on A, with \(\Lambda_x=\sum_a\mathcal I_{a\mid x}\) trace preserving, define the unnormalized conditional state of B as
\begin{equation}
\s_{a\mid x}=\Tr_A[(\mathcal I_{a\mid x}\otimes\id_B)(\rho_{AB})].
\end{equation}
\begin{theorem}[Unconditioned local operations do not signal]
For every bipartite state and each such local instrument,
\begin{equation}
\sum_a \s_{a\mid x}=\rho_B\qquad\text{for all }x.
\end{equation}
\end{theorem}
\begin{proof}
For every Bob effect \(F\), the dual of \(\Lambda_x\) obeys \(\Lambda_x^*(\I_A)=\I_A\). Consequently
\begin{align*}
\Tr\big[F\Tr_A((\Lambda_x\otimes\id_B)(\rho_{AB}))\big]
&=\Tr[(\Lambda_x^*(\I_A)\otimes F)\rho_{AB}]\\
&=\Tr[(\I_A\otimes F)\rho_{AB}]=\Tr(F\rho_B).
\end{align*}
Equality for every \(F\) proves the claim.
\end{proof}
Conditional states can depend on \(x,a\); their average cannot. Importantly, operational no-signalling is \emph{not} by itself the ontological assertion of no remote physical influence or disturbance.

\section{What Bell's theorem excludes}
Let \(x,y\in\{0,1\}\) and \(a,b\in\{-1,+1\}\). A measurement-independent Bell-local model factors as
\begin{equation}
 p(a,b\mid x,y)=\int d\lambda\,\mu(\lambda)\,p_A(a\mid x,\lambda)p_B(b\mid y,\lambda),
\label{eq:factorization}
\end{equation}
where \(\mu\) does not depend on \((x,y)\).
\begin{theorem}[CHSH bound]
Every model of the form \eqref{eq:factorization} satisfies
\begin{equation}
|S|=|E_{00}+E_{01}+E_{10}-E_{11}|\le2,
\qquad E_{xy}=\sum_{a,b}ab\,p(a,b\mid x,y).
\end{equation}
\end{theorem}
\begin{proof}
Define \(A_x(\lambda)=\sum_a a\,p_A(a\mid x,\lambda)\) and \(B_y(\lambda)=\sum_b b\,p_B(b\mid y,\lambda)\), each in \([-1,1]\). Pointwise,
\[
|A_0(B_0+B_1)+A_1(B_0-B_1)|\le |B_0+B_1|+|B_0-B_1|\le2.
\]
Integrating against \(\mu\) yields the bound.
\end{proof}
For \(|\Phi^+\rangle\), the choices \(A_0=Z,A_1=X\), and \(B_0=(Z+X)/\sqrt2,B_1=(Z-X)/\sqrt2\) give
\begin{equation}
E_{00}=E_{01}=E_{10}=\frac{1}{\sqrt2},\quad E_{11}=-\frac1{\sqrt2},\quad S=2\sqrt2.
\end{equation}
Thus these quantum predictions exclude \emph{the conjunction} of Bell-local factorization and measurement independence, not arbitrary deterministic descriptions. The quantum maximum in this setting is Tsirelson's bound \cite{Tsirelson1980}; operational randomness certification has additional explicit independence and device assumptions \cite{Pironio2010}.

\section{Deterministic response representations are not deterministic ontologies}
\begin{theorem}[Setting-independent-seed representation of finite tables]
Let \(\mathcal X,\mathcal Y,\mathcal O_A,\mathcal O_B\) be finite sets and \(p(a,b\mid x,y)\) any normalized conditional table. There is a random variable \(U\sim\operatorname{Uniform}[0,1]\), distributed independently of \((x,y)\), and deterministic functions
\begin{equation}
a=f(x,y,U),\qquad b=g(x,y,U),
\end{equation}
such that
\begin{equation}
\Prob(f(x,y,U)=a,\,g(x,y,U)=b)=p(a,b\mid x,y)
\end{equation}
for every allowed setting pair.
\end{theorem}
\begin{proof}
For each \((x,y)\), enumerate the outcome pairs by \(k=1,\ldots,N\) and set \(q_k^{xy}=p(a_k,b_k\mid x,y)\). Partition \([0,1)\) into consecutive half-open intervals \(I_k^{xy}\) of lengths \(q_k^{xy}\). Given \(U=u\), output \((a_k,b_k)\) for the unique interval containing \(u\). Zero-length intervals contribute no outcomes. Uniformity of \(U\) reproduces exactly every \(q_k^{xy}\); the distribution of the seed does not depend on settings.
\end{proof}

\begin{remark}[Precisely what is and is not deterministic]
The outcomes are determined \emph{conditional on the value of} \(U\). The theorem neither derives that value from deterministic initial conditions nor supplies a dynamical law for the preparation, measurement choices, or evolution of a physical system. One could draw the seed stochastically at preparation; nothing in the construction settles whether its origin is intrinsic randomness or ignorance of a richer state. Moreover \(f\) and \(g\) can both depend on the distant setting, so no Bell locality or relativistically acceptable microscopic implementation is implied. The theorem establishes \emph{representational non-uniqueness relative to the observed table}, not a deterministic completion of quantum mechanics.
\end{remark}

\begin{example}[A PR box: no-signalling but nonquantum]
Let \(x,y,a,b\in\{0,1\}\), put \(r(U)=0\) for \(U<1/2\) and \(r(U)=1\) otherwise, and define
\begin{equation}
a=r(U),\qquad b=r(U)\oplus(xy),
\end{equation}
where \(\oplus\) is addition modulo 2. The outcomes satisfy \(a\oplus b=xy\) with probability one and both local marginals equal \(1/2\), independently of the distant setting. With \(\pm1\) outputs \(A=(-1)^a\), \(B=(-1)^b\), the four correlations are \(E_{00}=E_{01}=E_{10}=1\) and \(E_{11}=-1\), hence
\begin{equation}
|S|=4>2\sqrt2.
\end{equation}
This is a Popescu--Rohrlich correlation \cite{PopescuRohrlich1994}. It obeys operational no-signalling but exceeds the quantum Tsirelson bound. Its seed representation therefore cannot establish physical realizability under quantum mechanics, still less a viable deterministic ontology.
\end{example}

\subsection{What changes when response functions are local?}
If instead the model requires \(a=f(x,U)\), \(b=g(y,U)\), the random variable \(U\) induces a joint distribution on the four setting-indexed outcomes \((A_0,A_1,B_0,B_1)\). All observed pair distributions are then marginals of a \emph{single} global distribution, and the CHSH restrictions follow. In the standard \(2\!\times\!2\!\times\!2\) Bell scenario, Fine's theorem establishes, subject to compatible one-party marginals, equivalence between existence of such a joint distribution and the complete relevant family of Bell inequalities \cite{Fine1982}. The unrestricted construction above provides no such setting-local outcomes. Thus there is no contradiction between it, Fine's theorem, and observed CHSH violations.

\begin{corollary}[Limited conclusion about ontological inference]
A finite conditional probability table, including a no-signalling table, does not by itself exclude all deterministic \emph{response-function representations} whose outputs may depend on both settings. It does not follow that every such table admits a deterministic \emph{physical theory} satisfying specified locality, dynamics, or measurement constraints.
\end{corollary}
\begin{proof}
The existence claim follows from Theorem 6.1. The physical-theory conclusion does not follow: the response functions are mathematical assignments, and the PR-box example shows that this representational class contains nonquantum tables.
\end{proof}

\section{The EPR implication and a proposed substitution}
\subsection{Logical non-equivalence and a two-observable counterexample}
Schnabel's revised argument \cite{Schnabel2026} distinguishes an implication [I], according to which certain prediction without disturbance, together with the absence of a theoretical counterpart for the predicted value, establishes incompleteness, from a proposed replacement [II], according to which such prediction excludes the value's genuinely random origin. Schnabel argues that the latter is an equivalent restatement of the former. We distinguish a logical claim about a theory's \emph{counterparts} from a physical claim about the \emph{causal generation} of an outcome. The distinction also requires a temporal specification of ``random''.

Let $Q$ be a set of target observables, with value set $V_q$ for
$q\in Q$. A \emph{prediction context} $c$ specifies the preparation,
Alice's measurement choice and registered record, and the space--time
conditions under which a value for Bob is inferred. Let $\mathcal D$
be the set of admissible triples $(q,v,c)$, where $v\in V_q$. Define
\begin{align*}
\predict(q,v\mid c)&:\ \text{in context $c$, Bob's value $v$ of $q$ is predicted with certainty},\\
\nodist(c)&:\ \text{the operations in $c$ do not physically disturb Bob},\\
\reality(q,v\mid c)&:\ \text{$v$ is attributed as an EPR element of reality for $q$ in $c$},\\
\represent_T(q,v\mid c)&:\ \text{theory $T$ represents this individual value attribution},\\
\intrinsic(q,v\mid c)&:\ \text{the attributed value originated through an objective chance event}.
\end{align*}
These predicates refer to an \emph{individual predicted value}, not merely
an operator or its probability distribution. The context index on
$\reality$ records the basis for attribution; it does not, without an
additional premise, assert that different counterfactual contexts can be
combined into one context-independent assignment. The last predicate
concerns stochastic generation at some stage, not necessarily a new
random draw when Bob is subsequently measured.

In this explicitly context-relative formalization, the sufficient EPR
reality criterion and the corresponding completeness definition are
\begin{align}
\mathbf{EPR_R}&:\quad
 \forall_{(q,v,c)\in\mathcal D}
 [\predict(q,v\mid c)\land\nodist(c)
 \Rightarrow\reality(q,v\mid c)],
 \label{eq:reality}\\
\comp(T)&\quad\Longleftrightarrow\quad
 \forall_{(q,v,c)\in\mathcal D}
 [\reality(q,v\mid c)\Rightarrow\represent_T(q,v\mid c)].
 \label{eq:completeness}
\end{align}
Here $\comp(T)$ means completeness relative to these EPR-attributed
individual values. The incompleteness implication and the
stochastic-origin reading of the proposed substitute become
\begin{align}
\mathbf I_T&:\ \forall_{(q,v,c)\in\mathcal D}
 [\predict(q,v\mid c)\land\nodist(c)\land\neg\represent_T(q,v\mid c)
 \Rightarrow\neg\comp(T)],\label{eq:IT}\\
\mathbf{II}_{\rm src}&:\ \forall_{(q,v,c)\in\mathcal D}
 [\predict(q,v\mid c)\land\nodist(c)
 \Rightarrow\neg\intrinsic(q,v\mid c)].\label{eq:II}
\end{align}
Equation~\eqref{eq:II} concerns \emph{stochastic origin}, not every
possible meaning of ``randomness at measurement.''

\begin{proposition}[Context- and value-indexed non-entailment]
Assume $\mathbf{EPR_R}$ and definition~\eqref{eq:completeness}.
Then $\mathbf I_T$ follows, but $\mathbf{II}_{\rm src}$ does not.
The non-entailment concerns the \emph{stochastic-origin reading} of [II]
and the explicitly postulated EPR reality and context-relative completeness
premises; it is not a refutation of every interpretation of Schnabel's [II].
The witness has two perfectly correlated alternative settings but is not a
model of the entire quantum EPR experiment or of its trusted-state assemblage.
\end{proposition}
\begin{proof}
For any admissible triple $(q,v,c)$, if
$\predict(q,v\mid c)\land\nodist(c)\land
\neg\represent_T(q,v\mid c)$ holds, the reality premise yields
$\reality(q,v\mid c)$. The universal completeness condition is then
false, proving $\mathbf I_T$.

To disprove entailment of $\mathbf{II}_{\rm src}$, consider a
\emph{local stochastic-source model} with settings
$x,y\in\{X,Z\}$. At preparation, an objectively stochastic source
generates independent fair signs $U_X,U_Z\in\{-1,+1\}$; after
preparation the realized signs are fixed and copied to two separated
stations. Each station reports the sign corresponding to its local
setting:
\begin{equation}
A_x=U_x,\qquad B_y=U_y,\qquad
p(a,b\mid x,y)=
\begin{cases}
\frac12\,\mathbf 1_{\{a=b\}},&x=y,\\[2pt]
\frac14,&x\ne y.
\end{cases}
\label{eq:two-context}
\end{equation}
The model is Bell-local and setting-independent. It yields
\begin{equation}
\Prob(B_j=A_j\mid x=y=j)=1,\quad
\Prob(B_j=\pm1\mid y=j)=\frac12,\quad j\in\{X,Z\},
\label{eq:two-context-marginals}
\end{equation}
and Alice's observation does not alter Bob's records. For any
$j\in\{X,Z\}$ and realized Alice outcome $a\in\{-1,+1\}$,
let $c_j^a$ denote the preparation and Alice's choice $j$ with record
$A_j=a$. Then
$\predict(j,a\mid c_j^a)$ and $\nodist(c_j^a)$ hold, so EPR's
criterion assigns $\reality(j,a\mid c_j^a)$. Yet
$\intrinsic(j,a\mid c_j^a)$ is true under the model's stipulated
objectively stochastic preparation. Hence $\mathbf{II}_{\rm src}$
is false for these admissible contexts.

Take an operational description $T_{\rm op}$ specifying only
Eq.~\eqref{eq:two-context}, without a representation of each realized
individual sign in the EPR sharp-value sense. Thus
$\represent_{T_{\rm op}}(j,a\mid c_j^a)$ is false by construction;
$\comp(T_{\rm op})$ is false by Eq.~\eqref{eq:completeness},
consistent with $\mathbf I_{T_{\rm op}}$. The countermodel is therefore an interpretation of the stated
\emph{logical} premises in which the specified stochastic-origin implication
fails. The objective nature of the initial chance event is an ontological
\emph{stipulation} of this countermodel, not a fact inferred from its
probability table; the same table has deterministic-seed representations.
\end{proof}

\noindent\textbf{Precise domain of the result.}
The proposition proves that $\mathbf{EPR_R}$ and
Eq.~\eqref{eq:completeness}, even supplemented with the stipulated local
no-disturbance of the shared-record example, do not logically entail
$\mathbf{II}_{\rm src}$. It does \emph{not} establish that Schnabel's
collision experiment has objective-chance preparation, that its outcomes
are pre-existing, or that a measurement-time stochastic theory is
inconsistent. A physical evaluation of the latter issues requires a
specified dynamics and measurement model. In particular, the
countermodel's matching of four $X/Z$ distributions does not yield a
local-hidden-\emph{quantum}-state explanation, as Proposition~\ref{prop:steering-two-setting}
shows.

\begin{remark}[Physical scope and the two meanings of ``random'']
The pairwise table in Eq.~\eqref{eq:two-context} coincides with the quantum statistics of \(\lvert\Phi^+\rangle\) restricted to local Pauli \(X\) and \(Z\) measurements, including the uncorrelated cross-basis outcomes. The local source model does \emph{not} reproduce the full set of Bell-test settings and cannot evade the CHSH exclusion of Bell-local theories. Objective stochastic preparation is a coherent physical \emph{model assumption}, not an experimental demonstration that nature uses this source.

If ``truly random'' instead means \emph{newly generated during a measurement interaction}, let $R_{\rm meas}(q,v\mid c)$ denote that different claim. The stochastic-source model above does \emph{not} establish $R_{\rm meas}(j,a\mid c_j^a)$: its outcomes were fixed at preparation. It therefore cannot serve as a physical counterexample to this stronger, measurement-time reading of [II]. Conversely, the identity $H(B_j\mid A_j,x=y=j)=0$ expresses perfect \emph{conditional} predictability; neither its truth nor the nonzero marginal entropy $H(B_j)=1$ determines \emph{when} the outcome became definite or whether an ontic chance event occurred. Conditioning on Alice's outcome is not a causal intervention on B, and Alice's record need not be locally available at B. These probabilities alone settle neither $R_{\rm meas}(j,a\mid c_j^a)$ nor its negation.

\textbf{Direct assessment of the measurement-time claim in Schnabel v2.} In the discussion following his uncertainty relation, Schnabel \cite{Schnabel2026} argues that a particle lacks a precise position and momentum before measurement and \emph{therefore} that its measured value must be genuinely random. The unstated implication is
\begin{equation}
\Delta_\rho x\,\Delta_\rho p\geq\frac{\hbar}{2}
\quad\nRightarrow\quad
\text{intrinsic stochasticity of an individual measurement outcome}.
\label{eq:not-intrinsic}
\end{equation}
The left-hand side constrains the variances of quantum observables \emph{in a state}; it is not an ontological theorem about when measurement results come into existence. For example, Bohmian mechanics gives definite particle configurations while recovering quantum position statistics, including their uncertainty constraints. Its particle momentum must not simply be equated with a sharp value of the canonical momentum observable (see Sec.~8). This suffices to block the inference that quantum uncertainty \emph{alone} excludes all deterministic descriptions. Similarly, producing an entangled state by a collision establishes a preparation procedure and quantum correlations, but does not supply the missing ontological premise. Schnabel explicitly presents completeness of quantum theory as an assumption in his physical discussion \cite{Schnabel2026}; it cannot also function as an independently proved consequence of that discussion.

Thus the relevant critique has \emph{two distinct parts}: Proposition~7.1 disproves the asserted logical substitution for the specified stochastic-origin reading, while the above counterargument identifies the unsupported inference from quantum uncertainty and entangling preparation to \emph{measurement-time} intrinsic randomness. Neither argument proves EPR's no-disturbance premise, and neither alone establishes a fully specified physical alternative to standard quantum theory.

The original EPR criterion is a \emph{sufficient assumption} concerning reality and non-disturbance. One may challenge it, as Schnabel intends, but doing so requires an independent argument. In particular, replacing ``element of reality with no theoretical counterpart'' by ``outcome not produced stochastically'' presupposes a further bridge such as \(\reality(q,v\mid c)\Rightarrow\neg\intrinsic(q,v\mid c)\), which is not part of Eqs.~\eqref{eq:reality}--\eqref{eq:completeness} and is false in the stochastic-source model. The proposition proves non-entailment for the stated reading of [II]; it does \emph{not} purport to demonstrate EPR's physical premises or to refute every possible alternative interpretation of Schnabel's argument.
\end{remark}

\noindent\textbf{Counterfactual context identification.}
EPR's original two-observable challenge considers \emph{alternative}
measurements on Alice's side applied to the same target system B. The
context-relative criterion in Eq.~\eqref{eq:reality}, by itself,
assigns no context-free joint values to incompatible observables. To
infer simultaneous elements of reality for one individual, one must
add an identification premise: attributions obtained under the
counterfactual Alice contexts are taken to refer to the same
undisturbed physical condition of B and to survive the alternative
choice. EPR motivates this step by its locality/no-disturbance
assumption, but it is \emph{not} an algebraic consequence of the
conditional probability table. The classical source in
Eq.~\eqref{eq:two-context} happens to admit joint values
$(U_X,U_Z)$; this model-specific fact cannot be inferred for the
trusted-qubit Bell-state assemblage, as the next proposition shows.

\noindent\textbf{Prior work.}
The inequality below is an elementary two-Pauli specialization of
established linear steering criteria, not a new steering inequality
\cite{Cavalcanti2009,Wiseman2007,Uola2020}.

\begin{proposition}[Two-setting steering bound for a trusted qubit]
\label{prop:steering-two-setting}
Let Alice's two measurements, labelled $j\in\{X,Z\}$, have outcomes
$a\in\{-1,+1\}$. Suppose Bob's corresponding measurements are
\emph{trusted qubit} measurements of the Pauli operators $\sigma_x$ and
$\sigma_z$. For the observed joint distributions define
\begin{equation}
 C_j:=\sum_{a,b=\pm1}ab\,p(a,b\mid j,j)
      =\sum_{a=\pm1}a\,\Tr(\sigma_{a\mid j}\sigma_j),
 \qquad j\in\{X,Z\},
 \label{eq:steering-correlators}
\end{equation}
where $\sigma_{a\mid j}$ is Bob's unnormalized conditional qubit state.
Every local-hidden-state (LHS) model satisfies the two-setting steering
inequality
\begin{equation}
 \boxed{C_X+C_Z\leq\sqrt{2}}.
 \label{eq:two-setting-steering}
\end{equation}
For $|\Phi^+\rangle$, Alice's and Bob's matched Pauli measurements give
$C_X=C_Z=1$, so $C_X+C_Z=2>\sqrt{2}$; the Bell-state assemblage is
therefore incompatible with an LHS model for these measurements.
\end{proposition}
\begin{proof}
An LHS decomposition has the form
$\sigma_{a\mid j}=\int d\lambda\,\mu(\lambda)
 p_A(a\mid j,\lambda)\rho_\lambda$,
where each $\rho_\lambda$ is a normalized qubit density operator. Put
$\alpha_j(\lambda)=\sum_{a=\pm1}a\,p_A(a\mid j,\lambda)$ and
$r_j(\lambda)=\Tr(\rho_\lambda\sigma_j)$; hence
$|\alpha_j(\lambda)|\leq1$. The Bloch-vector constraint gives
$r_x(\lambda)^2+r_z(\lambda)^2\leq1$. Consequently,
\begin{align}
 C_X+C_Z
 &=\int d\lambda\,\mu(\lambda)\bigl[
 \alpha_X(\lambda)r_x(\lambda)+
 \alpha_Z(\lambda)r_z(\lambda)\bigr]\\
 &\leq\int d\lambda\,\mu(\lambda)
 \bigl(|r_x(\lambda)|+|r_z(\lambda)|\bigr)\\
 &\leq\sqrt{2}\int d\lambda\,\mu(\lambda)
 \sqrt{r_x(\lambda)^2+r_z(\lambda)^2}\leq\sqrt{2}.
\end{align}
For $|\Phi^+\rangle$, the identities
$\langle\sigma_x\otimes\sigma_x\rangle=
\langle\sigma_z\otimes\sigma_z\rangle=1$
prove the asserted violation.
\end{proof}

\noindent\textbf{Operational qualification.}
The classical source in Eq.~\eqref{eq:two-context} indeed reproduces
\emph{all four} $X/Z$ joint-outcome distributions, and it is Bell-local
when the two stations are treated as unrestricted classical measurement
devices. Proposition~\ref{prop:steering-two-setting} does not contradict
that fact: its LHS inequality additionally requires that Bob's trusted
measurements act on a \emph{single qubit} described by valid density
operators. No such qubit-state ensemble can encode both independently
preassigned sharp signs $U_X$ and $U_Z$ while yielding the observed
matched correlations. Thus equality of a restricted classical
probability table is not equality of quantum conditional-state
assemblages. In particular, violation of the steering bound should not
be confused with violation of a Bell inequality, and it does not
invalidate the limited logical countermodel of Proposition~7.1
\cite{Wiseman2007,Uola2020,Cavalcanti2009}.

\subsection{Why correlated random outcomes do not settle EPR completeness}
Let \(J\) be a fair binary variable and let \(A=B=J\). Then
\begin{equation}
\Prob(B=b\mid A=b)=1,\qquad \Prob(B=b)=\frac12\qquad(b=0,1).
\end{equation}
The example disproves the \emph{probabilistic} assertion that a variable with perfect conditional predictability must have a degenerate marginal distribution. It says nothing without further assumptions about the ontological origin of \(J\), the physical interpretation of non-disturbance, or the adequacy of a theory's counterparts for \(B\). A stochastic \emph{preparation} of \(J\) can yield a value fixed before either measurement, illustrating that ``stochastic origin'' and ``unpredictable at the moment of measurement'' are also distinct notions. In particular, a full EPR analysis cannot be replaced by a one-observable correlation alone.

\section{Does an uncertainty relation prove intrinsic randomness?}
Schnabel's revision \cite{Schnabel2026} also appeals to the generation of position--momentum entanglement by an elastic collision \cite{SchnabelCollision2025}, and infers from a local uncertainty relation that neither particle can have a precise position or momentum before measurement and that a particular measurement value must therefore be truly random. The two steps require separate evaluation.

For a quantum state \(\rho\), Robertson's relation for a canonical pair is
\begin{equation}
\Delta_\rho x\,\Delta_\rho p\ge\frac{\hbar}{2},\qquad
(\Delta_\rho A)^2=\Tr(\rho A^2)-\Tr(\rho A)^2.
\label{eq:heisenberg}
\end{equation}
The variances refer to the outcome distributions of the state as represented by the quantum formalism. Relation \eqref{eq:heisenberg} excludes states with simultaneously vanishing \emph{quantum variances} for both observables, not all ontic models in which additional precise variables exist. In particular, Bohmian theories retain the ordinary quantum predictions while providing definite particle configurations and nonlocal deterministic dynamics \cite{BohmI,BohmII}. These theories do not generally identify an ontic particle momentum with a sharp value of the quantum momentum observable; that distinction matters when interpreting Eq.~\eqref{eq:heisenberg}.

\begin{proposition}[An uncertainty bound does not by itself establish ontic indeterminism]
The state-dependent inequality \eqref{eq:heisenberg}, together with correct conditional EPR correlations, is insufficient to conclude that all admissible microscopic descriptions assign intrinsically random origins to individual outcomes.
\end{proposition}
\begin{proof}
The premises constrain the quantum distributions of outcomes. They are respected by nonlocal deterministic interpretations such as the Bohmian framework \cite{BohmI,BohmII}. Existence of such an interpretation suffices to refute a logical inference from those quantum-statistical premises alone to the exclusion of every deterministic ontology. This is a counterexample to the \emph{unrestricted inference}, not a proof that Bohmian mechanics is uniquely correct.
\end{proof}

An explicit physical process that prepares an entangled state is important for feasibility, but preparation dynamics does not supply the missing inference from quantum variances to an ontic claim. Similarly, distinguishing \emph{inference after a correlated measurement} from \emph{physical creation of a remote value} requires a separate model of locality or disturbance. No-signalling in Theorem~4.2 does not resolve that ontological question.

\subsection{Continuous-variable steering criterion for the collision proposal}
Proposition~\ref{prop:steering-two-setting} is a two-setting
\emph{discrete-variable} statement for trusted Pauli observables on
Bob's qubit. Schnabel's elastic-collision proposal instead concerns
continuous-variable position--momentum correlations
\cite{SchnabelCollision2025,SchnabelCollisionCorrection2025}. Neither
the classical two-bit countermodel nor its qubit steering violation
calculates the conditional variances for that collision-generated
state. We therefore apply an explicit idealized collision map below; the actual
finite-duration model requires additional physical inputs.

For example, with $x_B,p_B$ the canonical observables on the trusted
Bob system, let $a$ denote Alice's outcome in a position-inference
setting and $a'$ her outcome in a momentum-inference setting. Define
\begin{align}
 \Delta_{\rm inf}^2 x_B
 &=\int da\,p(a\mid x_A)\,
   \operatorname{Var}(x_B\mid a,x_A),\\
 \Delta_{\rm inf}^2 p_B
 &=\int da'\,p(a'\mid p_A)\,
   \operatorname{Var}(p_B\mid a',p_A).
 \label{eq:cv-inferred}
\end{align}
These are average conditional variances, corresponding to optimal conditional-mean estimators. For finite second moments, trusted canonical observables, and an
LHS decomposition with a setting-independent ensemble of valid
quantum states for B, the local uncertainty relation plus convexity
of conditional variance imply the Reid-type bound
\begin{equation}
 \Delta_{\rm inf} x_B\,\Delta_{\rm inf} p_B\geq \frac{\hbar}{2}.
 \label{eq:cv-reid}
\end{equation}
Indeed, under an LHS model each inferred variance is at least the
ensemble average of the corresponding variance in $\rho_\lambda$;
Cauchy--Schwarz and
$\Delta_{\rho_\lambda}x_B\,\Delta_{\rho_\lambda}p_B
\geq\hbar/2$ then give Eq.~\eqref{eq:cv-reid}
\cite{Reid1989,Cavalcanti2009}. A strict violation witnesses continuous-variable steering in this
trusted-Bob scenario, not the ontological origin of individual
outcomes. The following calculation evaluates an \emph{explicit idealized}
instantaneous-collision Gaussian map compatible with the stated
mass and preparation assumptions. It is \emph{not} an exact quantum
dynamical reconstruction of the corrected experimental proposal: the
published analysis uses a semiclassical treatment and supplies neither
a finite-duration scattering unitary nor complete noise parameters
\cite{SchnabelCollision2025,SchnabelCollisionCorrection2025}.

\subsection{Finite-squeezing inferred variances for an idealized elastic-collision map}
\label{sec:collision-cov}
The corrected collision article specifies independent initially Gaussian
wave packets, a one-dimensional elastic encounter, and a nominal mass
ratio $m_B/m_A=3$ \cite{SchnabelCollision2025,SchnabelCollisionCorrection2025}.
Related kinematic-entanglement conditions for unequal-mass wave packets were already studied by Schulman \cite{Schulman2004} and by Harshman and Hutton \cite{HarshmanHutton2008}; the latter also separates kinematic and dynamical effects in one-dimensional scattering. The calculation here is a steering-oriented Gaussian benchmark, \emph{not} a new discovery of the known separability condition.
The following quantitative result concerns only the \emph{stipulated
instantaneous, perfectly reflecting, one-dimensional collision map}.
In Proposition~\ref{prop:collision-cov} the inputs are additionally
independent, pure, unchirped, minimum-uncertainty Gaussians, and
measurements take place at the output reference plane without further
propagation. These are independent modelling assumptions, not
consequences of energy--momentum conservation or parameters inferred
from the proposed trapped-ion Coulomb encounter. In particular, the
map is \emph{not} obtained here by solving the apparatus Hamiltonian.

Write $\kappa=m_B/m_A>0$, $d=1+\kappa$, and center all quadratures
on their preparation means. The canonical phase-space map associated
with classical elastic velocity reversal in the relative coordinate is
\begin{equation}
\begin{pmatrix}p_A'\\p_B'\end{pmatrix}
=P_\kappa\begin{pmatrix}p_A\\p_B\end{pmatrix},\quad
P_\kappa=\frac1d
\begin{pmatrix}1-\kappa&2\\2\kappa&\kappa-1\end{pmatrix},
\qquad
\begin{pmatrix}x_A'\\x_B'\end{pmatrix}
=P_\kappa^{\mathsf T}\begin{pmatrix}x_A\\x_B\end{pmatrix}.
\label{eq:collision-map}
\end{equation}
Here $P_\kappa^2=I$; the position map is its inverse transpose and
therefore preserves the canonical commutators. This mathematical
consistency establishes neither a scattering phase nor the probability
of reflection for finite-width ion wave packets. For $\kappa=3$,
\begin{equation}
(p_A',p_B')=\tfrac12(-p_A+p_B,3p_A+p_B),\quad
(x_A',x_B')=\tfrac12(-x_A+3x_B,x_A+x_B),
\label{eq:collision-3}
\end{equation}
consistent with the corrected article's leading $-1/2,+1/2$
coefficients of the initial $x_A$ fluctuations. A statement about the
real apparatus at a fixed measurement time additionally requires its
actual scattering dynamics and propagation.

\begin{proposition}[Ideal Gaussian elastic-collision covariance benchmark]
\label{prop:collision-cov}
Suppose the two incoming packets are independent, pure,
minimum-uncertainty Gaussians at the collision reference plane, with
\begin{equation}
\operatorname{Var}(x_A)=A>0,\quad
\operatorname{Var}(x_B)=B>0,\quad
\operatorname{Var}(p_A)=u=\frac{\hbar^2}{4A},\quad
\operatorname{Var}(p_B)=v=\frac{\hbar^2}{4B},
\label{eq:initial-widths}
\end{equation}
and vanishing initial $x$--$p$ covariances. For the Gaussian output
of Eq.~\eqref{eq:collision-map}, measurements immediately after that
map have optimal conditional-mean inferred variances
\begin{align}
\Delta_{\mathrm{inf}}^2 x_B
 &=\frac{d^2 AB}{(\kappa-1)^2 A+4\kappa^2 B},
 \label{eq:collision-vx}\\
\Delta_{\mathrm{inf}}^2 p_B
 &=\frac{d^2 uv}{(\kappa-1)^2u+4v}
 =\frac{d^2\hbar^2}{4[(\kappa-1)^2B+4A]}.
 \label{eq:collision-vp}
\end{align}
If $r=A/B$, their dimensionless Reid product obeys
\begin{equation}
\mathcal R(\kappa,r)
 :=\frac{2\Delta_{\mathrm{inf}}x_B
              \Delta_{\mathrm{inf}}p_B}{\hbar}
 =\frac{(1+\kappa)^2\sqrt r}
 {\sqrt{[(\kappa-1)^2r+4\kappa^2]
         [(\kappa-1)^2+4r]}}.
\label{eq:collision-ratio}
\end{equation}
For $\kappa\ne1$, the criterion $\mathcal R<1$ is satisfied
except when $r=\kappa$. For $\kappa=1$ or $r=\kappa$,
$\mathcal R=1$ and this witness gives no violation. These
statements are exact \emph{for the specified pure-Gaussian output of
Eq.~\eqref{eq:collision-map} at the reference plane}; they are not
predictions for an unspecified finite-duration interaction.
\end{proposition}
\begin{proof}
For the position and momentum covariance matrices of the incoming
product state, let $V_x=\operatorname{diag}(A,B)$ and
$V_p=\operatorname{diag}(u,v)$. Equation~\eqref{eq:collision-map}
gives $V_x'=P_\kappa^{\mathsf T}V_xP_\kappa$ and
$V_p'=P_\kappa V_pP_\kappa^{\mathsf T}$.
Because $|\det P_\kappa|=1$, both transformed determinants are
$AB$ and $uv$, respectively. For any bivariate Gaussian vector,
$\operatorname{Var}(Y\mid X)=\det(V_{XY})/V_{XX}$; here
\begin{equation}
(V_x')_{AA}=\frac{(\kappa-1)^2A+4\kappa^2 B}{d^2},
\quad
(V_p')_{AA}=\frac{(\kappa-1)^2u+4v}{d^2}.
\end{equation}
The Schur complements yield Eqs.~\eqref{eq:collision-vx}--
\eqref{eq:collision-vp}. In addition, direct expansion gives
\begin{equation}
\big[(\kappa-1)^2r+4\kappa^2\big]
\big[(\kappa-1)^2+4r\big]-(1+\kappa)^4r
=4(\kappa-1)^2(r-\kappa)^2\ge0,
\label{eq:collision-identity}
\end{equation}
which proves the stated equality and strict-violation cases.
\end{proof}

For the nominal ratio $\kappa=3$ this simplifies to
\begin{equation}
\boxed{\Delta_{\mathrm{inf}}^2x_B=\frac{4AB}{A+9B},\quad
\Delta_{\mathrm{inf}}^2p_B=\frac{\hbar^2}{A+B},\quad
\mathcal R(3,r)=\frac{4\sqrt r}{\sqrt{(r+9)(r+1)}}.}
\label{eq:collision-three}
\end{equation}
The equality point $r=3$ is significant: the output $x$ and $p$
interparticle cross-covariances both vanish (in this unchirped Gaussian
model), so the nominal $3{:}1$ mass ratio
\emph{alone} does not guarantee entanglement or steering for arbitrary
incoming widths. For a strongly momentum-squeezed A and strongly
position-squeezed B ($r\gg3$), $\mathcal R\simeq4/\sqrt r$.
If the packets were initially harmonic-trap ground states with
frequencies $\Omega_A$ and $\Omega_B$, neglecting pre-collision
spreading gives $r=\kappa\Omega_B/\Omega_A$. For an
\emph{illustrative, not experimentally specified} ratio
$\Omega_B/\Omega_A=10$ with $\kappa=3$, one obtains $r=30$ and
$\mathcal R\approx0.6301$, i.e.
$\Delta_{\rm inf}x_B\Delta_{\rm inf}p_B\approx0.3150\hbar$.
The corrected article proposes a potassium--caesium pair with
$\kappa\approx133/39$, not exactly three; the general formulas
above permit that ratio to be used without treating it as $3$.

\noindent\textbf{Relation to earlier collision-purity work.}
Schulman identified a mass--width matching condition for the absence
of kinematic entanglement, and Harshman--Hutton analyzed purity and
scattering entanglement for unequal masses
\cite{Schulman2004,HarshmanHutton2008}. In the present notation,
the \emph{unchirped, pure} input separability condition is
$m_A A=m_B B$, i.e. $r=\kappa$. There is also a useful identity:
if $V_A^{\rm out}$ is the $2\times2$ covariance of Alice's
$(x_A',p_A')$ and $\rho_A$ her reduced Gaussian state, then
\begin{equation}
\mathcal R(\kappa,r)
 =\frac{\hbar}{2\sqrt{\det V_A^{\rm out}}}
 =\operatorname{Tr}(\rho_A^2).
\label{eq:purity-identity}
\end{equation}
This follows because the global output is a pure two-mode Gaussian,
$\det V_{AB}^{\rm out}=(\hbar/2)^4$, and the position and momentum
blocks have no mutual $x$--$p$ covariance at the reference plane.
Equation~\eqref{eq:purity-identity} is an identity \emph{specific to
this pure Gaussian model} (and the indicated axes), not a general
identification of a Reid product with state purity. In particular,
chirp and thermal mixing require the analysis below.

\noindent\textbf{Free-flight and dynamical qualification.}
At a further free-flight time $\tau$ after the ideal collision,
and still assuming no initial $x$--$p$ covariance at the chosen
reference plane, let
$M^{-1}=\operatorname{diag}(1/m_A,1/m_B)$. Then
\begin{equation}
V_x(\tau)=V_x'+\tau^2 M^{-1}V_p'M^{-1},\qquad
V_p(\tau)=V_p',
\label{eq:collision-flight}
\end{equation}
and the inferred position variance becomes
\begin{equation}
\Delta_{\rm inf}^2x_B(\tau)
=[V_x(\tau)]_{BB}-\frac{[V_x(\tau)]_{AB}^2}
{[V_x(\tau)]_{AA}}.
\label{eq:collision-flight-infer}
\end{equation}
A finite flight time, initial Gaussian chirp, unequal collision times,
trapping forces, wave-packet overlap or a nontrivial scattering phase
may modify this \emph{fixed-axis} benchmark. The corrected article does
not specify a complete finite-duration scattering channel and noise
budget; hence Eqs.~\eqref{eq:collision-ratio} and
\eqref{eq:collision-flight-infer} are conditional, not apparatus-level
predictions. The difference between witness degradation and loss of
entanglement or steerability is quantified next.

\subsection{Pre-collision spreading, mixed inputs, and complete Gaussian covariance}
\label{sec:precollision-robust}
A minimal extension retains the ideal collision map but propagates the
\emph{full} position--momentum covariance of each incident packet.
Use the ordering $z=(x_A,x_B,p_A,p_B)^{\mathsf T}$ and the
symmetrized covariance $\Gamma_{ij}=\langle\{\Delta z_i,
\Delta z_j\}\rangle/2$. With independent initial particles, write
\begin{equation}
\Gamma_{\rm in}=
\begin{pmatrix}X&C\\ C^{\mathsf T}&Q\end{pmatrix},\quad
X=\operatorname{diag}(A_A,A_B),\quad
Q=\operatorname{diag}(u_A,u_B),\quad
C=\operatorname{diag}(c_A,c_B).
\label{eq:gamma-in}
\end{equation}
Each physical input obeys $A_i u_i-c_i^2\ge\hbar^2/4$.
The enlarged computation allows arbitrary $c_i$, not merely the
unchirped minimum-uncertainty inputs of Proposition~\ref{prop:collision-cov}.
The ideal collision is the symplectic map
\begin{equation}
S_{\kappa}=\begin{pmatrix}P_\kappa^{\mathsf T}&0\\0&P_\kappa\end{pmatrix},
\qquad \Gamma_{\rm out}=S_\kappa\Gamma_{\rm in}S_\kappa^{\mathsf T},
\label{eq:gamma-collision}
\end{equation}
where $P_\kappa$ is defined in Eq.~\eqref{eq:collision-map}.
For fixed local $x$ and $p$ measurements at the output, the directly
tested Reid parameter becomes
\begin{align}
V_{x|x_A}&=\Gamma_{x_Bx_B}-
\frac{\Gamma_{x_Ax_B}^2}{\Gamma_{x_Ax_A}},&
V_{p|p_A}&=\Gamma_{p_Bp_B}-
\frac{\Gamma_{p_Ap_B}^2}{\Gamma_{p_Ap_A}},\\
\mathcal R_{\rm axes}&=\frac{2\sqrt{V_{x|x_A}V_{p|p_A}}}{\hbar},
\label{eq:axes-witness}
\end{align}
where all entries refer to $\Gamma_{\rm out}$ (or to its subsequent
propagation). This formula is exact for the stipulated Gaussian
covariance model even with input chirp or mixing.

For example, take $A_i^{(0)}>0$ as the position variance of a
\emph{reference pure} squeezed state for particle $i\in\{A,B\}$.
Let the actual initial Gaussian covariance be $\nu_i$ times this
pure reference covariance, with $\nu_i\ge1$; its initial position
and momentum variances are $\nu_i A_i^{(0)}$ and
$\nu_i\hbar^2/(4A_i^{(0)})$, respectively.
If the particle evolves freely for $t_i$ before the reference collision,
\begin{equation}
\begin{aligned}
A_i(t_i)&=\nu_i\left[A_i^{(0)}+
\frac{\hbar^2 t_i^2}{4m_i^2 A_i^{(0)}}\right],\\
u_i(t_i)&=\nu_i\frac{\hbar^2}{4A_i^{(0)}},\\
c_i(t_i)&=\nu_i\frac{\hbar^2t_i}{4m_i A_i^{(0)}}.
\end{aligned}
\label{eq:incoming-spread}
\end{equation}
Here the momentum variance $u_i(t_i)$ is time independent, while
$\nu_i$ multiplies both quadrature variances and the covariance,
preserving $\det V_i=\nu_i^2\hbar^2/4$.
An initially pure packet ($\nu_i=1$) has the characteristic spreading
time $t_i^*=2m_i A_i^{(0)}/\hbar$. Neglecting pre-collision
free spreading is a controlled approximation only when
$|t_i|/t_i^*\ll1$ for each strongly squeezed species.

The experimental proposal includes a new trap after preparation.
The free-propagation approximation above is not the unique possible
incident dynamics. In a locally harmonic potential of frequency
$\Omega_i$ the corresponding one-mode symplectic propagator is
\begin{equation}
F_i^{\Omega}(t)=
\begin{pmatrix}
\cos(\Omega_i t)&\sin(\Omega_i t)/(m_i\Omega_i)\\
-m_i\Omega_i\sin(\Omega_i t)&\cos(\Omega_i t)
\end{pmatrix};
\label{eq:harmonic-pre}
\end{equation}
$F_i^\Omega V_i^{(0)}(F_i^\Omega)^{\mathsf T}$ should replace the
free result when the uncoupled harmonic approximation is justified.
The always-present inter-ion force and a finite collision duration
would instead require a coupled dynamical treatment.

\subsection{Fixed-axis Reid witnesses versus optimized Gaussian steering}
\label{sec:gauss-opt}
For a physical two-mode Gaussian output, reorder the covariance into
$(x_A,p_A,x_B,p_B)$ and partition it as
\begin{equation}
\Gamma^{AB}=\begin{pmatrix}V_A&K\\K^{\mathsf T}&V_B\end{pmatrix},
\quad M_{B|A}=V_B-K^{\mathsf T}V_A^{-1}K,
\quad \mathcal R_G=\frac{2\sqrt{\det M_{B|A}}}{\hbar}.
\label{eq:gauss-schur}
\end{equation}
With a single trusted Gaussian mode at B, $\mathcal R_G<1$ is the
necessary and sufficient covariance condition for steering from A to B
by Gaussian measurements \cite{Kogias2015}. This condition allows
locally adapted measurement quadratures. A nonviolation of the
\emph{fixed-axis} witness in Eq.~\eqref{eq:axes-witness} therefore
does not establish absence of steering, and a local post-collision
unitary does not itself change bipartite entanglement or steerability.

For a \emph{pure} two-mode Gaussian state, $\det\Gamma^{AB}=(\hbar/2)^4$.
The block determinant identity yields
\begin{equation}
\boxed{\mathcal R_G
 =\frac{\hbar}{2\sqrt{\det V_A}}
 =\operatorname{Tr}(\rho_A^2).}
\label{eq:gauss-purity}
\end{equation}
Thus the Gaussian-steering parameter equals reduced purity for any
pure two-mode Gaussian output. For the unchirped reference-plane
state, $\mathcal R_{\rm axes}=\mathcal R_G=\mathcal R(\kappa,r)$;
chirp or subsequent free evolution generally breaks the equality
between the two \emph{witnesses}, not between pure-state steering and
entanglement.

One particularly instructive case is a common free-evolution time
$t_A=t_B=t$ before the ideal collision. Define
$D=\operatorname{diag}(1/m_A,1/m_B)$. The identity
$P_\kappa^{\mathsf T}D=DP_\kappa$ implies that the collision map
commutes with equal-time local free propagation. Hence the final state
is related to the $t=0$ output by a product of local free evolutions,
so $\mathcal R_G$ and its entanglement are unchanged even when the
fixed-axis Reid test degrades. For the illustrative \emph{pure} input
$\hbar=m_A=1$, $m_B=3$, $A_A^{(0)}=30$, $A_B^{(0)}=1$ and
$t_B^*=6$, direct covariance propagation gives
\begin{center}
\begin{tabular}{@{}ccc@{}}
\toprule
$t/t_B^*$ & $\mathcal R_{\rm axes}$ & $\mathcal R_G$ \\
\midrule
$0$   & $0.6301$ & $0.6301$ \\
$0.5$ & $0.6852$ & $0.6301$ \\
$1$   & $0.8047$ & $0.6301$ \\
$2$   & $1.0279$ & $0.6301$ \\
\bottomrule
\end{tabular}
\end{center}
The final row loses this fixed-setting Reid violation but does
\emph{not} lose Gaussian steering \emph{within the stipulated
pure-state collision model}. The numbers are dimensionless
illustrations, not predictions for trapped ions. The invariance of
$\mathcal R_G$ here follows from equal-time free propagation and the
specific ideal map; unequal incident propagation times, other traps,
or a nontrivial scattering unitary need not preserve it.

\subsection{Noise robustness, realistic mass ratio, and dynamical limits}
\label{sec:ion-noise}
To assess state-preparation impurity, take both incoming covariance
matrices to be uniformly inflated by the same factor $\nu\ge1$
relative to the pure inputs while keeping their squeezing axes fixed:
$\Gamma_{\rm in}^{(\nu)}=\nu\Gamma_{\rm in}^{(1)}$.
Linearity of Eq.~\eqref{eq:gamma-collision} and of the conditional
Schur complements gives
\begin{equation}
\mathcal R_{\rm axes}^{(\nu)}=\nu\mathcal R_{\rm axes}^{(1)},
\qquad \mathcal R_G^{(\nu)}=\nu\mathcal R_G^{(1)}.
\label{eq:thermal-scaling}
\end{equation}
For the unchirped $\kappa=3$, $r=30$ example,
$\mathcal R^{(1)}\simeq0.6301$, so this Gaussian steering
violation requires $\nu<1/0.6301\simeq1.5871$. If the factor is
interpreted as a common squeezed-thermal occupation,
$\nu=2\bar n+1$, the corresponding bound is $\bar n<0.2935$.
These are \emph{model-conditional uniform covariance-inflation}
thresholds, not an experimentally established noise budget or a
requirement on an actual trap temperature. The conversion
$\nu=2\bar n+1$ uses the assumed squeezed-thermal parametrization;
unequal $\nu_A,\nu_B$ are handled by the full covariance formulas
\eqref{eq:gamma-in}--\eqref{eq:gauss-schur}. A Reid nonviolation
does not by itself imply that the mixed output is separable or unsteerable
under a more general measurement class.

Independent additive measurement-readout noise must not be conflated
with mixing of the physical quantum state. Let $n_{qA},n_{qB}\ge0$
be readout-noise variances for $q=x,p$, with no correlated noise.
Then the experimentally accessible fixed-axis inference variance is
\begin{equation}
V_{q|q_A}^{\rm obs}
=\Gamma_{q_Bq_B}+n_{qB}
-\frac{\Gamma_{q_Aq_B}^2}{\Gamma_{q_Aq_A}+n_{qA}},
\qquad q\in\{x,p\}.
\label{eq:readout}
\end{equation}
In particular, with ideal Alice readout and equal \emph{relative}
added noise on Bob's two conditional variances,
$n_{xB}=\eta V_{x|x_A}$ and $n_{pB}=\eta V_{p|p_A}$,
$\mathcal R_{\rm axes}^{\rm obs}=(1+\eta)\mathcal R_{\rm axes}$;
for $\kappa=3,r=30$ a violation persists only if
$\eta<0.5871$. Correlated detector errors, calibration offsets,
and apparatus-specific efficiencies require separate modelling.

For the suggested potassium--caesium isotopes the mass-number
approximation yields
\begin{equation}
\kappa_{\rm KCs}\simeq\frac{133}{39}=3.410256\ldots,\qquad
\frac{p_B'}{p_A'}\simeq\frac{\kappa_{\rm KCs}-1}{2}=1.205128\ldots
\quad(p_A^{\rm in}\simeq0).
\label{eq:isotopes}
\end{equation}
Thus the exactly equal outgoing momenta in the ideal $3{:}1$ example
are \emph{not} the corresponding relation for the isotope pair.
The optimal regression gains at finite input widths are
$g_x=(V_x')_{AB}/(V_x')_{AA}$ and
$g_p=(V_p')_{AB}/(V_p')_{AA}$; they must not be replaced by the
limit in Eq.~\eqref{eq:isotopes}. If the \emph{illustrative}, not
experimentally established ground-state trap-frequency ratio is
$\Omega_B/\Omega_A=10$, then at the unspread reference plane
$r=\kappa_{\rm KCs}\Omega_B/\Omega_A\simeq34.103$ and
$\mathcal R\simeq0.60895$; the corresponding finite-width gains are
$g_x\simeq-0.6048$ and $g_p\simeq1.0403$. Neither the trap ratio
nor the resulting steering value is claimed for the actual apparatus.

Finally, the instantaneous, perfectly reflecting canonical map is an
\emph{assumption} about the scattering channel, not a consequence of
energy and momentum conservation alone. To approximate a finite
quantum collision, one must justify a reflected mode with effectively
unit amplitude over the occupied momentum band, and control the
momentum dependence of its scattering phase (notably quadratic phase
terms that create chirp), the collision duration relative to relevant
trap and spreading scales, residual transmitted channels, and
inter-ion forces during the encounter. The corrected paper specifies
an implementation concept but does not fix all parameters of such a
unitary and detector model \cite{SchnabelCollision2025,SchnabelCollisionCorrection2025}.
Accordingly, these are \emph{conditional analytic benchmarks},
not independently validated predictions for the proposed apparatus.
Their practical applicability can be tested only after establishing
the relevant incident state, scattering channel, propagation, and
readout model; they neither demonstrate steering in the proposed
experiment nor decide whether individual outcomes have intrinsic
ontological randomness.

\subsection{Scope of the collision calculation and requirements for an apparatus comparison}
\label{sec:collision-scope}
The mathematical results in Sections~\ref{sec:collision-cov}--\ref{sec:ion-noise}
refer to the model defined by Eq.~\eqref{eq:collision-map}, with
additional assumptions specified for each calculation. To prevent
conclusions from exceeding those assumptions, the warranted statements
are as follows.
\begin{enumerate}[label=(\roman*),leftmargin=2.0em]
\item \textbf{Ideal kinematics.} The map $S_\kappa$ is symplectic and
  defines a Gaussian transformation. This does not establish that a
  finite-duration Coulomb scattering process realizes that map over
  the occupied momentum range.
\item \textbf{Pure, unchirped reference-plane inputs.} Equations
  \eqref{eq:collision-vx}--\eqref{eq:collision-ratio} give the
  conditional variances and a Reid violation for the specified input
  state and immediate output measurements. They do not show that the
  suggested ion preparation satisfies these input covariances.
\item \textbf{Spreading and traps.} Equations~\eqref{eq:gamma-in}--
  \eqref{eq:harmonic-pre} give covariance propagation only when the
  assumed free or uncoupled harmonic dynamics applies. They are not a
  solution of the coupled, finite-duration collision Hamiltonian.
\item \textbf{Steering witnesses.} A value $\mathcal R_{\rm axes}<1$
  witnesses steering for a state obeying the trusted-Bob assumptions
  when the specified quadratures are measured; $\mathcal R_G<1$ is
  the Gaussian-measurement steering condition for a specified
  two-mode Gaussian state. Neither inequality establishes that the
  proposed hardware produces that state, and neither establishes
  a microscopic stochastic origin of measurement outcomes.
\item \textbf{Illustrative parameters.} The illustrative trap-frequency ratio
  $\Omega_B/\Omega_A=10$ and common-flight times are chosen for
  sensitivity tests; $\nu$ and $\eta$ independently parametrize
  stipulated noise models. None of these quantities is an apparatus
  measurement in this paper. The isotope
  mass-number ratio is an approximation, not an experimental
  determination of output covariance or detector performance.
\end{enumerate}
Before a claim about a specific apparatus is warranted one would need
(a) independently justified incident covariance matrices at the
interaction plane, including any chirp and impurity; (b) a scattering
unitary or quantitatively validated approximation incorporating the
interaction potential, collision duration and momentum-dependent phase;
(c) the propagation and trapping dynamics from scattering to detection;
and (d) calibrated measurement settings, efficiencies and noise.
Only then can the measured or predicted output covariance be compared
with the Reid or Gaussian-steering thresholds. No such complete
apparatus-level calculation or experimental verification is claimed here.

\section{Radioactive memorylessness does not establish irreducible indeterminism}
\begin{proposition}[Predetermined exponential lifetimes are statistically memoryless]
Let lifetimes be sampled at preparation from \(T\sim\mathrm{Exp}(\lambda)\) with \(\lambda>0\), and suppose an individual system's assigned value of \(T\) is then fixed. For \(s,t\ge0\),
\begin{equation}
\Prob(T>t+s\mid T>t)=e^{-\lambda s}.
\end{equation}
Thus exponential memorylessness alone cannot exclude a model with individually predetermined lifetimes.
\end{proposition}
\begin{proof}
By the definition of conditional probability,
\[
\Prob(T>t+s\mid T>t)=\frac{\Prob(T>t+s)}{\Prob(T>t)}
=\frac{e^{-\lambda(t+s)}}{e^{-\lambda t}}=e^{-\lambda s}.
\]
\end{proof}

This is a counterexample to a \emph{particular inference from survival statistics}, not a microscopic model of nuclear decay. The initial sampling of \(T\) may itself be stochastic; ``predetermined event time after preparation'' must not be conflated with ``deterministic preparation and fundamental dynamics.'' Absence of a statistical ageing effect also does not imply absence of internal time evolution: a constant hazard rate and physical dynamics are not mutually exclusive.

Furthermore, a perfectly exponential survival probability is not a universal exact prediction of closed-system quantum decay. For an initially pure state of finite energy variance,
\begin{equation}
P_{\mathrm{surv}}(t)=|\langle\psi_0|e^{-iHt/\hbar}|\psi_0\rangle|^2
=1-\frac{(\Delta H)^2}{\hbar^2}t^2+o(t^2),\qquad t\to0,
\end{equation}
under the usual differentiability assumptions. Hence the initial derivative is zero, unlike that of \(e^{-\lambda t}\) for \(\lambda>0\). Nonexponential regimes and related quantum-Zeno considerations are standard \cite{MisraSudarshan1977,Chiu1977}. These facts constrain realistic decay models independently of the elementary memorylessness counterexample.

An alpha-particle detection can also indicate a correlated daughter nucleus, but this single-observable correlation does not furnish EPR's alternative position/momentum measurement structure. It is therefore not, on its own, a refutation of the full EPR inference.

\section{Relation to quantum steering}
The assemblage \(\{\s_{a\mid x}\}_{a,x}\) of Eq.~(4.3) contains Bob's unnormalized conditional states. It has a local-hidden-state (LHS) model if
\begin{equation}
\s_{a\mid x}=\int d\lambda\,\mu(\lambda)\,p(a\mid x,\lambda)\rho_\lambda,
\end{equation}
where \(\rho_\lambda\) is a normalized quantum state of B independent of Alice's setting. Failure to admit such a model witnesses steering in the stated scenario \cite{Wiseman2007,Uola2020}. No-signalling does not imply an LHS decomposition, and the unrestricted functions of Theorem~6.1 must \emph{not} be mistaken for an LHS model. Proposition~\ref{prop:steering-two-setting} gives an explicit two-setting witness for this distinction: the trusted-qubit Bell-state correlations violate Eq.~\eqref{eq:two-setting-steering} although a classical shared-bit model reproduces their restricted joint-outcome table.

Our earlier analysis \cite{SienickiEPR} distinguished context-indexed conditional assignments from simultaneous, context-free values. The present contribution places that distinction beside the formal non-entailment result of Proposition~7.1 and the explicit PR-box limitation of seed representations. The existence of a particular representation must never be confused with an independently specified ontology of the experiment.

\section{Discussion: what a valid resolution would have to prove}
A defensible resolution of the EPR issue must identify which of the following questions it answers:
\begin{enumerate}
\item \textbf{Consistency:} quantum mechanics consistently predicts non-degenerate local marginals, perfect conditional correlations and incompatible measurement settings (Theorem~4.1).
\item \textbf{Operational causality:} local unconditioned quantum operations do not enable signalling, without automatically establishing every ontological no-disturbance premise (Theorem~4.2).
\item \textbf{Bell-locality:} measurement-independent factorizable models obey CHSH; the relevant quantum predictions violate it (Theorem~5.1).
\item \textbf{Representation versus ontology:} a deterministic response-function simulation of a table is not a physical dynamics. The PR box makes this limitation especially clear (Section~6).
\item \textbf{Logical EPR criterion and steering:} the value- and context-indexed incompleteness implication follows from an assumed reality criterion and completeness definition, but does not entail a claim about stochastic origin of outcomes (Proposition~7.1). Counterfactual attributions require an additional cross-context identification premise. Reproducing a restricted classical table does not reproduce the Bell-state assemblage when Bob's qubit measurements are trusted (Proposition~\ref{prop:steering-two-setting}); the stipulated ideal Gaussian collision map yields explicit inferred variances (Proposition~\ref{prop:collision-cov}), while Sections~\ref{sec:precollision-robust}--\ref{sec:ion-noise} examine spreading, assumed noise models, and Gaussian steering. Section~\ref{sec:collision-scope} states why these calculations do not establish steering for a specific finite-duration apparatus.
\item \textbf{Intrinsic randomness:} quantum uncertainty relations and a specified preparation mechanism do not, by themselves, exclude every ontologically deterministic account (Section~8).
\item \textbf{Decay analogy:} memoryless survival alone cannot distinguish predetermined from intrinsically random event times, and a single decay correlation does not duplicate the EPR alternative-observable construction (Section~9).
\end{enumerate}
A proof of unrestricted ontological completeness would require an independently justified definition of admissible completion together with a theorem excluding \emph{all} members of that class. Bell tests constrain local causal structures very strongly, but their implications have to be stated with the relevant assumptions. Conversely, a hidden-variable representation without dynamics, locality, and operational predictions is not a proposed alternative physical theory.

\section{Conclusion}
Quantum mechanics demonstrably accommodates perfect EPR-type correlations and locally non-degenerate statistics without algebraic inconsistency. Bell inequalities exclude an important class of local models, not every possible deterministic interpretation. Any finite conditional probability table admits a deterministic response-function representation with a setting-independent seed when both local and distant settings may enter the responses; such a representation can reproduce even superquantum no-signalling statistics and therefore carries no automatic ontological significance.

The established two-setting qubit steering bound demonstrates that the Bell-state assemblage cannot be replaced by the restricted classical shared-bit model once Bob's Pauli measurements are trusted. Under the additional assumptions of an instantaneous reflecting symplectic map and specified pure Gaussian inputs, the collision benchmark yields a Reid violation except at the identified equality cases. Its covariance extensions describe how \emph{stipulated} propagation, covariance-inflation and readout-noise models alter the witnesses; some fixed-axis violations disappear even when optimized Gaussian steering persists within that ideal model. None of the numerical examples establishes the incoming state, scattering transformation, or steering performance of the proposed potassium--caesium apparatus, for which a validated quantum-dynamical and detector model is still needed. The key logical point is sharper than a probability example: EPR's sufficient reality criterion and its context- and value-dependent completeness definition yield an implication about missing theoretical counterparts, with any cross-context identification of elements of reality as a separate physical premise. They do not, without an additional linking premise, yield Schnabel's proposed implication about the intrinsic origin of measurement outcomes. Nor does the quantum uncertainty relation alone provide that premise. The correct conclusion is an assumption-sensitive delimitation of EPR claims, not an unrestricted proof of quantum completeness or a newly discovered inconsistency in the 1935 argument.

\bigskip
\noindent\textbf{Statements and declarations.} This manuscript contains theoretical analysis only, no new experimental data, and no fitting or validation against measurements from the proposed apparatus. The Bell, CHSH, no-signalling, seed-simulation, and exponential-memorylessness techniques are established results; no priority is claimed for them. The restricted stochastic-source countermodel clarifies the logical scope of one asserted equivalence; it is not an empirically established account of EPR experiments. The collision calculations assume an instantaneous ideal-reflection map and specified Gaussian inputs, with independently chosen examples of propagation, covariance inflation and detector noise. The prior collision-entanglement literature is explicitly acknowledged.\par\smallskip
\noindent\textbf{Computational reproducibility.} A self-contained Python/NumPy verification script is included as ancillary material (\texttt{anc/verify\_covariance.py}). It checks the symplectic map, Gaussian covariance calculations, and displayed numerical illustrations. The script tests the \emph{specified model only}; it does not simulate the finite-duration ion collision, verify its physical realizability, or use experimental data. No public code repository or experimental dataset is asserted.
\par\smallskip
\noindent\textbf{Declaration on the use of generative AI.} The authors used ChatGPT (OpenAI) to assist with literature and reference checks, language editing, \LaTeX{} preparation, and the development of computational verification scripts. The authors reviewed and revised AI-assisted material and take full responsibility for the manuscript's arguments, calculations, accuracy, references, and conclusions. AI-generated suggestions were not treated as independent scientific evidence or independent verification. The AI system is not an author.


\begin{thebibliography}{99}
\bibitem{EPR1935} A. Einstein, B. Podolsky, and N. Rosen, ``Can Quantum-Mechanical Description of Physical Reality Be Considered Complete?'', \emph{Phys. Rev.} \textbf{47}, 777--780 (1935). \href{https://doi.org/10.1103/PhysRev.47.777}{doi:10.1103/PhysRev.47.777}.
\bibitem{Bohr1935} N. Bohr, ``Can Quantum-Mechanical Description of Physical Reality be Considered Complete?'', \emph{Phys. Rev.} \textbf{48}, 696--702 (1935). \href{https://doi.org/10.1103/PhysRev.48.696}{doi:10.1103/PhysRev.48.696}.
\bibitem{Bell1964} J. S. Bell, ``On the Einstein Podolsky Rosen paradox'', \emph{Physics Physique Fizika} \textbf{1}, 195--200 (1964). \href{https://doi.org/10.1103/PhysicsPhysiqueFizika.1.195}{doi:10.1103/PhysicsPhysiqueFizika.1.195}.
\bibitem{CHSH1969} J. F. Clauser, M. A. Horne, A. Shimony, and R. A. Holt, ``Proposed Experiment to Test Local Hidden-Variable Theories'', \emph{Phys. Rev. Lett.} \textbf{23}, 880--884 (1969). \href{https://doi.org/10.1103/PhysRevLett.23.880}{doi:10.1103/PhysRevLett.23.880}.
\bibitem{BohmI} D. Bohm, ``A Suggested Interpretation of the Quantum Theory in Terms of `Hidden' Variables. I'', \emph{Phys. Rev.} \textbf{85}, 166--179 (1952). \href{https://doi.org/10.1103/PhysRev.85.166}{doi:10.1103/PhysRev.85.166}.
\bibitem{BohmII} D. Bohm, ``A Suggested Interpretation of the Quantum Theory in Terms of `Hidden' Variables. II'', \emph{Phys. Rev.} \textbf{85}, 180--193 (1952). \href{https://doi.org/10.1103/PhysRev.85.180}{doi:10.1103/PhysRev.85.180}.
\bibitem{Wiseman2007} H. M. Wiseman, S. J. Jones, and A. C. Doherty, ``Steering, Entanglement, Nonlocality, and the Einstein-Podolsky-Rosen Paradox'', \emph{Phys. Rev. Lett.} \textbf{98}, 140402 (2007). \href{https://doi.org/10.1103/PhysRevLett.98.140402}{doi:10.1103/PhysRevLett.98.140402}.
\bibitem{Reid1989} M. D. Reid, ``Demonstration of the Einstein-Podolsky-Rosen paradox using nondegenerate parametric amplification'', \emph{Phys. Rev. A} \textbf{40}, 913--923 (1989). \href{https://doi.org/10.1103/PhysRevA.40.913}{doi:10.1103/PhysRevA.40.913}.
\bibitem{Uola2020} R. Uola, A. C. S. Costa, H. C. Nguyen, and O. G\"uhne, ``Quantum steering'', \emph{Rev. Mod. Phys.} \textbf{92}, 015001 (2020). \href{https://doi.org/10.1103/RevModPhys.92.015001}{doi:10.1103/RevModPhys.92.015001}.
\bibitem{Pironio2010} S. Pironio, A. Ac\'{\i}n, S. Massar, \emph{et al.}, ``Random numbers certified by Bell's theorem'', \emph{Nature} \textbf{464}, 1021--1024 (2010). \href{https://doi.org/10.1038/nature09008}{doi:10.1038/nature09008}.
\bibitem{SienickiEPR} M. Sienicki and K. Sienicki, ``EPR Revisited: Context-Indexed Elements of Reality and Operational Completeness'', \href{https://arxiv.org/abs/2511.01930}{arXiv:2511.01930} (2025; revised 2026).
\bibitem{Schnabel2026} R. Schnabel, ``Discovery of the Solution to the `Einstein--Podolsky--Rosen Paradox' '', \href{https://arxiv.org/abs/2604.09826v2}{arXiv:2604.09826v2} (revised 7 November 2026).
\bibitem{SchnabelCollision2025} R. Schnabel, ``Discovery of entanglement generation by elastic collision to realise the original Einstein-Podolsky-Rosen thought experiment'', \emph{npj Quantum Information} \textbf{11}, 76 (2025), corrected version of record. \href{https://doi.org/10.1038/s41534-025-01028-7}{doi:10.1038/s41534-025-01028-7}.
\bibitem{DaviesLewis1970} E. B. Davies and J. T. Lewis, ``An operational approach to quantum probability'', \emph{Commun. Math. Phys.} \textbf{17}, 239--260 (1970). \href{https://doi.org/10.1007/BF01647093}{doi:10.1007/BF01647093}.
\bibitem{Fine1982} A. Fine, ``Hidden Variables, Joint Probability, and the Bell Inequalities'', \emph{Phys. Rev. Lett.} \textbf{48}, 291--295 (1982). \href{https://doi.org/10.1103/PhysRevLett.48.291}{doi:10.1103/PhysRevLett.48.291}.
\bibitem{Tsirelson1980} B. S. Cirel'son, ``Quantum generalizations of Bell's inequality'', \emph{Lett. Math. Phys.} \textbf{4}, 93--100 (1980). \href{https://doi.org/10.1007/BF00417500}{doi:10.1007/BF00417500}.
\bibitem{PopescuRohrlich1994} S. Popescu and D. Rohrlich, ``Quantum nonlocality as an axiom'', \emph{Found. Phys.} \textbf{24}, 379--385 (1994).
\bibitem{MisraSudarshan1977} B. Misra and E. C. G. Sudarshan, ``The Zeno's paradox in quantum theory'', \emph{J. Math. Phys.} \textbf{18}, 756--763 (1977). \href{https://doi.org/10.1063/1.523304}{doi:10.1063/1.523304}.
\bibitem{Chiu1977} C. B. Chiu, E. C. G. Sudarshan, and B. Misra, ``Time evolution of unstable quantum states and a resolution of Zeno's paradox'', \emph{Phys. Rev. D} \textbf{16}, 520--529 (1977). \href{https://doi.org/10.1103/PhysRevD.16.520}{doi:10.1103/PhysRevD.16.520}.
\bibitem{Cavalcanti2009} E. G. Cavalcanti, S. J. Jones, H. M. Wiseman, and M. D. Reid, ``Experimental criteria for steering and the Einstein-Podolsky-Rosen paradox'', \emph{Phys. Rev. A} \textbf{80}, 032112 (2009). \href{https://doi.org/10.1103/PhysRevA.80.032112}{doi:10.1103/PhysRevA.80.032112}.
\bibitem{Schulman2004} L. S. Schulman, ``Evolution of Wave-Packet Spread under Sequential Scattering of Particles of Unequal Mass'', \emph{Phys. Rev. Lett.} \textbf{92}, 210404 (2004). \href{https://doi.org/10.1103/PhysRevLett.92.210404}{doi:10.1103/PhysRevLett.92.210404}.
\bibitem{HarshmanHutton2008} N. L. Harshman and G. Hutton, ``Entanglement Generation in the Scattering of One-Dimensional Particles'', \emph{Phys. Rev. A} \textbf{77}, 042310 (2008). \href{https://doi.org/10.1103/PhysRevA.77.042310}{doi:10.1103/PhysRevA.77.042310}.
\bibitem{Kogias2015} I. Kogias, A. R. Lee, S. Ragy, and G. Adesso, ``Quantification of Gaussian Quantum Steering'', \emph{Phys. Rev. Lett.} \textbf{114}, 060403 (2015). \href{https://doi.org/10.1103/PhysRevLett.114.060403}{doi:10.1103/PhysRevLett.114.060403}.
\bibitem{SchnabelCollisionCorrection2025} R. Schnabel, ``Author Correction: Discovery of entanglement generation by elastic collision to realise the original Einstein-Podolsky-Rosen thought experiment'', \emph{npj Quantum Information} \textbf{11}, 130 (2025). \href{https://doi.org/10.1038/s41534-025-01081-2}{doi:10.1038/s41534-025-01081-2}.
\end{thebibliography}
\end{document}